\documentclass[12pt,preprint]{aastex}
\usepackage{graphicx,amssymb,amsmath,times}
\shorttitle{Neutrinos from GRBs}
\shortauthors{Taboada \& D'Agostino}
\slugcomment{\today}

\begin{document}

\title{Correlating prompt GRB photons with neutrinos}
\author{Ignacio Taboada and Michelangelo V. D'Agostino}
\affil{Physics Dept. University of California. Berkeley, CA 94720}
\email{itaboada@berkeley.edu}

\begin{abstract}
      It is standard in theoretical neutrino astrophysics to use a broken
      power law approximation, based on the Band function, to describe the
      average photon flux of the prompt emission of Gamma-Ray Bursts. We will 
      show that this approximation overestimates the contribution of high
      energy $\gamma$-rays (and underestimates low energy $\gamma$-rays). As a
      consequence models that rely on this approximation overestimate neutrino
      event rate by a factor of $\approx$~2 depending on Earth's column density
      in the direction of the GRB. Furthermore the characteristic energy of
      neutrinos that trigger a km$^3$ detector is typically 10$^{16}$~eV,
      higher than previously predicted. We also provide a new broken power law 
      approximation to the Band function and show that it properly represents
      the photon spectra.
\end{abstract}


\section{Introduction}

It is commonly believed that prompt emission by GRBs is due to synchrotron
radiation by electrons accelerated in internal shocks associated with
relativistic jets (with a bulk Lorentz boost $\Gamma$ of 100-1000). A review
of the theoretical and observational status of GRBs is beyond the scope of
this paper. The reader is referred to \citet{2006RPPh...69.2259M}. The average
prompt photon emission is often described by fitting a Band function 
\citep{1993ApJ...413..281B}. High energy neutrino emission by GRBs in
coincidence with the prompt $\gamma$-ray photons has been proposed as a
consequence of GRBs being a candidate to produce (UHE) ultra high energy (up
to $\approx$~10$^{20}$~eV) cosmic rays
\citep{1995PhRvL..75..386W,1995ApJ...453..883V}.

\citet{1997PhRvL..78.2292W} have calculated the diffuse flux due to
GRBs. Their work has been further improved by calculating neutrino emission
for individual bursts \citep{2004APh....20..429G} and by performing detailed
GEANT4 simulations of proton-photon interactions in the internal shocks
\citep{2006PhRvD..73f3002M}. This body of work supports the hypothesis that
neutrino detection by km$^3$ Cherenkov detectors will probably be in
coincidence with a handful of bright GRBs.

Previous studies have correlated photon emission with neutrino emission by
approximating the photon spectra with a broken power law. We will show that
this approximation overestimates the contribution of high energy photons (and
conversely underestimates the contribution from low energy photons). This has
consequences in the number of events expected by neutrino telescopes such as
IceCube \citep{2007..IceCube..ICRC} and KM3NET \citep{2007..KM3NET..Faro} and in the
characteristic energy of the neutrinos detected.

In section \S \ref{sec:band} we describe the Band function. In section \S
\ref{sec:photopion} we follow the standard calculation of HE neutrinos
emission from GRBs. In section \S \ref{sec:bandapprox} we compare the Band
function with its broken power law approximation. In section \S
\ref{sec:eventrates} we calculate neutrino event rates. And in section \S
\ref{sec:discussion} we discuss the consequences of the calculation shown
here.

\section{Prompt photon spectra}
\label{sec:band}

The time-averaged prompt GRB photon flux is often described by fitting
a Band function \citep{1993ApJ...413..281B}:
\begin{equation}
  \label{eqn:band}
  \frac{dN_\gamma}{dE_\gamma}^{\mathrm{Band}}(E_\gamma)
= A_\gamma \left\{
  \begin{array}{lr}
 \left(\frac{E_\gamma}{100 \mathrm{keV}}\right)^{\alpha_\gamma}
 e^\frac{-E_\gamma(2+{\alpha_\gamma})}{E_{\mathrm{pk}}}
 \;\; &  E_\gamma<\epsilon^b_\gamma\\
 \left[ \frac{(\alpha_\gamma - \beta_\gamma)}{2+\alpha_\gamma}
 \frac{E_{\mathrm{pk}}}{100 \mathrm{keV}} 
 \right]^{\alpha_\gamma-\beta_\gamma} e^{\beta_\gamma-\alpha_\gamma}
 \left(\frac{E_\gamma}{100 \mathrm{keV}}\right)^{\beta_\gamma} \;\; &  E_\gamma>\epsilon^b_\gamma
  \end{array}
\right. .
\end{equation}

The parameters of the Band function are the amplitude $A_\gamma$, the
(asymptotic) low-energy spectral index $\alpha_\gamma$, the high-energy
spectral index $\beta_\gamma$ and the peak energy $E_{\mathrm{pk}}$ of the
$\nu F_\nu$ distribution. Typical values for the parameters above are:
$\alpha_\gamma \approx -1$, $\beta_\gamma \approx-2.2$, $\epsilon^b_\gamma
\approx$~30~keV--1000~keV and $A_\gamma \approx$ 0.001--1
photons~cm$^{-2}$~s$^{-1}$~keV$^{-1}$. The break energy $\epsilon^b_\gamma$
and the peak energy $E_{\mathrm{pk}}$ are related by:
\begin{equation}
\epsilon^b_\gamma = \frac{\alpha_\gamma -
  \beta_\gamma}{2+\alpha_\gamma}E_{\mathrm{pk}} .
\end{equation}


\section{Photo-pion production of neutrinos}
\label{sec:photopion}

There are many good descriptions of the processes that lead to HE neutrinos
from prompt GRB photons in the literature. We will not repeat the full
derivation here. Instead we refer the reader to \citet{2004APh....20..429G}
and we only present here the major features relevant to this paper.

Protons accelerated in internal shocks interact with GRB photons via the
process:
\begin{equation}
\label{eqn:pgamma}
p+\gamma \rightarrow \Delta^+ \rightarrow \pi^+[+n] \rightarrow \mu^+ + \nu_\mu
 \rightarrow e^+ + \nu_\mu  + \bar{\nu}_{\mu} + \nu_{e}\;
\end{equation}
and
\begin{equation}
\label{eqn:pi0}
p+\gamma \rightarrow \Delta^+ \rightarrow \pi^0 + p\;.
\end{equation}

Because the photon-proton interaction has to create a $\Delta$ resonance, given
a photon energy there is a minimum proton energy for which
Eqs.\ref{eqn:pgamma} and \ref{eqn:pi0} can take place. In Earth's reference
frame:
\begin{equation}
E_p \ge \frac{1}{(1+z)^2}\frac{m^2_\Delta - m^2_p}{4 E_\gamma} \Gamma^2.
\end{equation}
Correspondingly the neutrinos resulting from Eq.\ref{eqn:pgamma} have a
minimum energy:
\begin{equation}
\label{eqn:nuphotoncorr}
E_\nu \ge \frac{1}{4} <x_{p \rightarrow \pi^+}>\frac{1}{(1+z)^2}\frac{m^2_\Delta - m^2_p}{4 E_\gamma} \Gamma^2,
\end{equation}
where $<x_{p \rightarrow \pi^+}> \approx 1/5$ is the fraction of the energy
transferred to the charged pion from the initial proton energy and the factor
of 1/4 arises because on average each one of the four final leptons in
Eq.\ref{eqn:pgamma} has the same energy.

It is customary to approximate the Band function fit to the average photon
spectra as a broken power law (which we label approximation \textit{A}):
\begin{equation}
  \label{eqn:bandapprox}
  \frac{dN_\gamma}{dE_\gamma}^{\mathrm{A}} = A_\gamma \left\{
  \begin{array}{lr}
    \left(\frac{E_\gamma}{100 keV}\right)^{\alpha_\gamma}  & E_\gamma < \epsilon^{\mathrm b}_\gamma \\
    \left(\frac{E_\gamma}{100 keV}\right)^{\beta_\gamma}   & E_\gamma \ge \epsilon^{\mathrm b}_\gamma
  \end{array}
\right. .
\end{equation}

Supposing that protons have a power law spectrum $dN_p/dE_p \sim E_p^{-2}$, the
neutrino spectrum traces Eq.\ref{eqn:bandapprox}:
\begin{equation}
  \label{eqn:wb}
  \frac{\mathrm{d}N_\nu}{\mathrm{d}E_\nu}^{A} = A_\nu \left\{
  \begin{array}{lr}
    ({E}/{\epsilon^{\mathrm b}_\nu})^{\alpha_\nu} & 
    E < \epsilon^{\mathrm b}_\nu \\ 
    ({E}/{\epsilon^{\mathrm b}_\nu})^{\beta_\nu} & 
    \epsilon^{\mathrm b}_\nu \le E \le \epsilon^s_\nu \\ 
    ({E}/{\epsilon^{\mathrm b}_\nu})^{\beta_\nu} ({E}/{\epsilon^s_\nu})^{-2}
    & E > \epsilon^s_\nu 
  \end{array}
\right. ,
\end{equation}
where $\alpha_\nu = -\beta_\gamma -3$, $\beta_\nu = -\alpha_\gamma -3$ and the 
neutrino break energy $\epsilon^{\mathrm b}_\nu$ is taken from the minimum
energy in Eq.\ref{eqn:nuphotoncorr}:
\begin{equation}
\epsilon^b_\nu = \frac{1}{20} \frac{1}{(1+z)^2}\frac{m^2_\Delta - m^2_p}{4
  \epsilon^b_\gamma} \Gamma^2.
\end{equation}
The spectrum is steeper above $\epsilon^s_\nu$ because of synchrotron energy
losses by charged pions. Typically $\epsilon^s_\nu \approx 10^{16}$~eV. Muons
also suffer from synchrotron losses, but following usual approximations we
ignore this effect. 

The neutrino flux normalization $A_\nu$ is obtained by supposing that the
neutrino fluence (ignoring synchrotron losses) is proportional to the
bolometric photon fluence, $S_\gamma$: 
\begin{equation}
\label{eqn:nunorm}
A_\nu \; \propto \;
\int^{E_\nu^{\mathrm{max}}}_{E_\nu^{\mathrm{min}}} E_\nu \frac{dN_\nu}{dE_\nu}
dE_\nu 
\; \propto \; 
S_\gamma = \int^\infty_0 dE_\gamma \; E_\gamma \frac{dN_\gamma}{dE_\gamma}.
\end{equation}

Here the choice of minimum neutrino energy, $E_\nu^{\mathrm{min}}$ is
unimportant and the maximum neutrino energy $E_\nu^{\mathrm{max}}$ can be set
so that the maximum proton energy is comparable to the highest energy cosmic
rays ($E_p^{\mathrm{max}} \sim 10^{20}$~eV \& $E_\nu^{\mathrm{max}} \sim
5\cdot10^{18}$~eV ). Depending on the spectral indices $\alpha_\nu$ and 
$\beta_\nu$, the neutrino normalization $A_\nu$ is independent or a very weak
function of $E_\nu^{\mathrm{max}}$.

It is possible to calculate the neutrino flux without approximating the Band
function. In this case the neutrino spectra is:
\begin{equation}
  \label{eqn:nu_band}
  \frac{\mathrm{d}N_\nu}{\mathrm{d}E_\nu}^{\mathrm{Band}} = A_\nu \left\{
  \begin{array}{lr}
    ({E_\nu}/{\epsilon^{\mathrm b}_\nu})^{\alpha_\nu} & 
    E_\nu < \epsilon^{\mathrm b}_\nu \\ 
    e^{-(\alpha_\nu-\beta_\nu)(\epsilon^{\mathrm b}_\nu/E_\nu -1)}
    ({E_\nu}/{\epsilon^{\mathrm b}_\nu})^{\beta_\nu} & 
    \epsilon^{\mathrm b}_\nu \le E_\nu \le \epsilon^s_\nu \\ 
    e^{-(\alpha_\nu-\beta_\nu)(\epsilon^{\mathrm b}_\nu/E_\nu -1)}
    ({E_\nu}/{\epsilon^{\mathrm b}_\nu})^{\beta_\nu} ({E_\nu}/{\epsilon^s_\nu})^{-2}
    & E_\nu > \epsilon^s_\nu 
  \end{array}
\right. .
\end{equation}
Here we have approximated the flux above the syncrhotron energy to a power
law. This approximation is correct as long as $\epsilon^s_\nu$ is sufficiently
larger than $\epsilon^{\mathrm b}_\nu$. The normalization, $A_\nu$, is
obtained using Eq.\ref{eqn:nunorm}.

\section{Approximations of the Band function}
\label{sec:bandapprox}

\begin{figure}
\begin{center}
\includegraphics[width=0.3\textwidth]{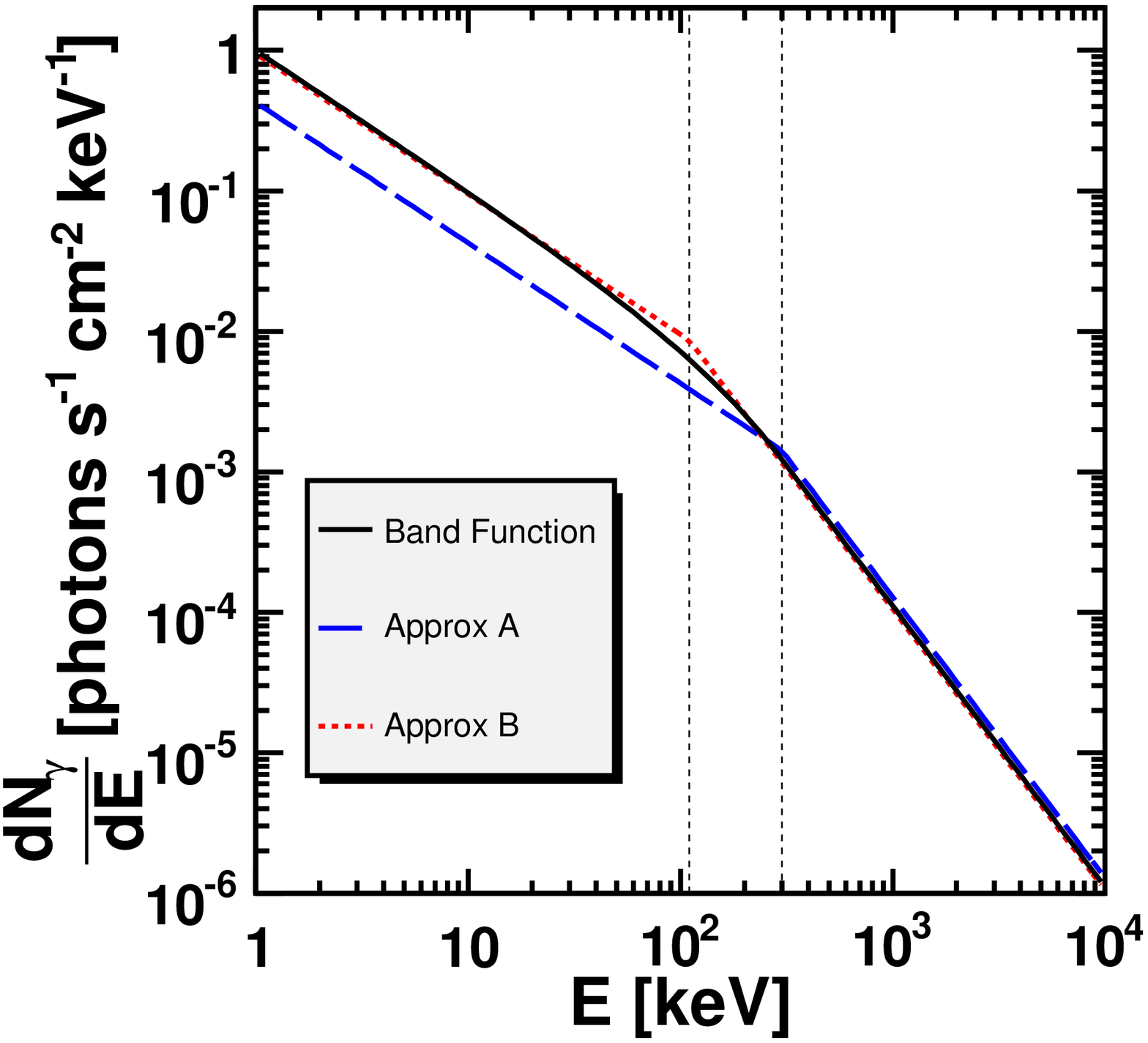}
\includegraphics[width=0.3\textwidth]{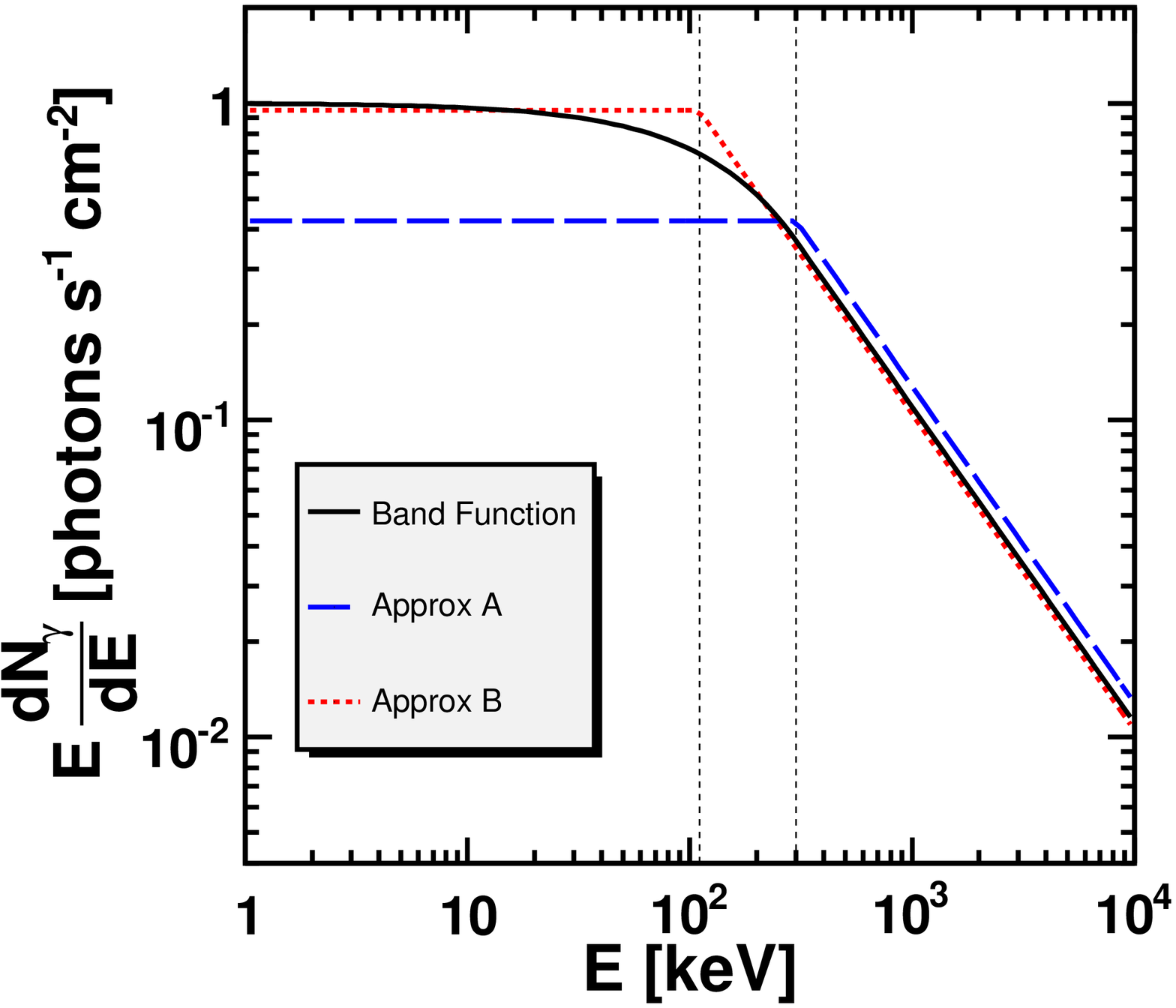}
\includegraphics[width=0.3\textwidth]{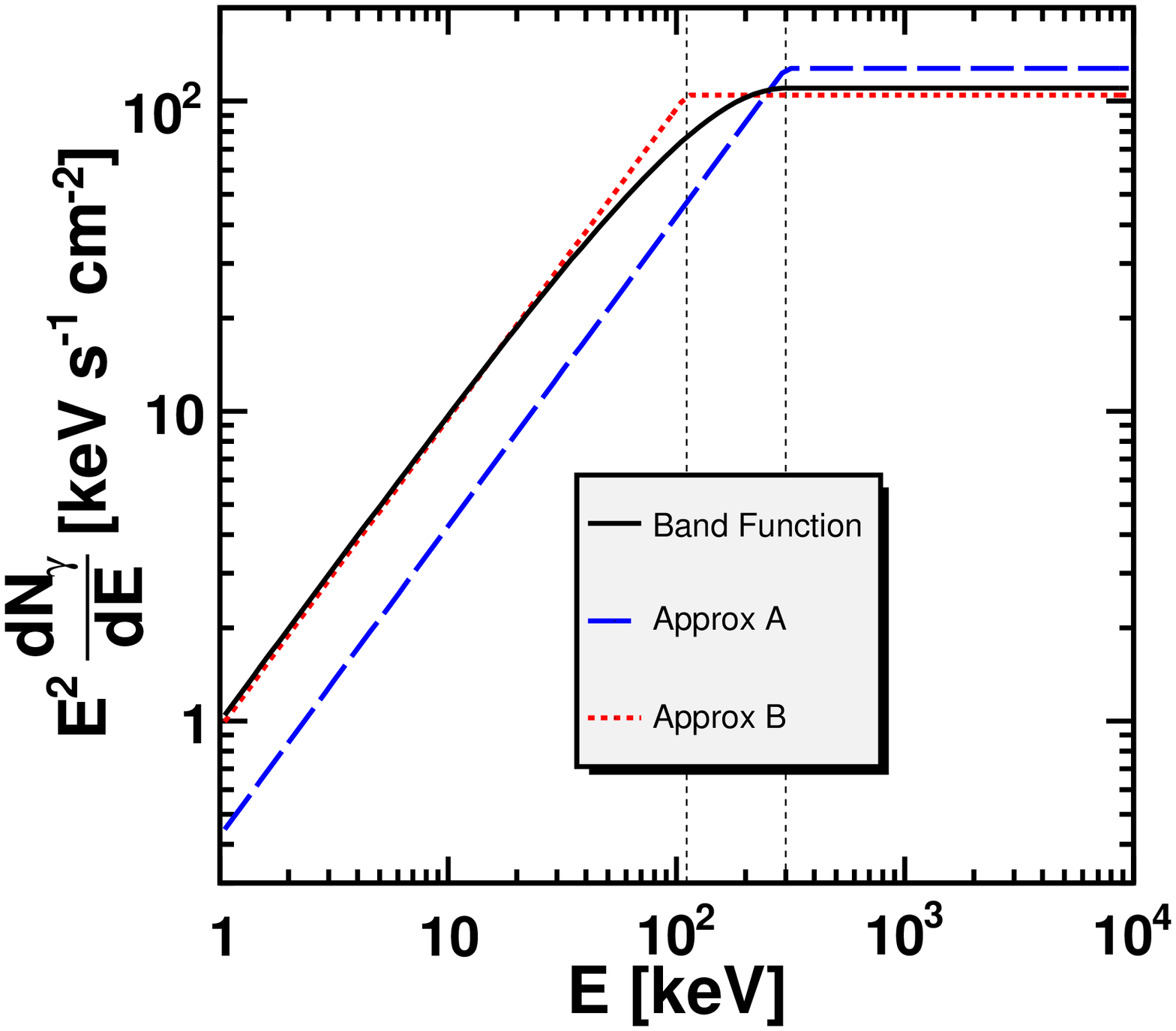}
\caption{\label{fig:bandA} The band function is compared to approximations A
  and B. The three plots show $dN_\gamma/dE_\gamma$, $E
  \mathrm{d}N_\gamma/\mathrm{d}E_\gamma$ and $E^2
  \mathrm{d}N_\gamma/\mathrm{d}E_\gamma$. The Integral of $E
  \mathrm{d}N_\gamma/\mathrm{d}E_\gamma$ (fluence) for approximations A and B
  has been normalized to the fluence of the band function. For this example we
  have chosen $A_\gamma = 10^{-2}$~photons~cm$^{-2}$~s$^{-1}$~keV$^{-1}$,
  $\alpha_\gamma = -1$, $\beta_\gamma = -2$ and $\epsilon^b_\gamma =
  300$~keV. For each plot, the two vertical lines indicate $\epsilon^b_\gamma$
  (right line) and $\bar{\epsilon}_\gamma$ (left line).}
\end{center}
\end{figure}

Figure \ref{fig:bandA} shows the Band function with the parameters:
$\alpha_\gamma = -1$, $\beta_\gamma = -2$, $\epsilon^b_\gamma = 300$~keV and
$A_\gamma = 0.01$~photons~cm$^{-2}$~s$^{-1}$~keV$^{-1}$. Also shown is the
corresponding approximation A from Eq.\ref{eqn:bandapprox}. The normalization
of approximation A has been chosen so that the bolometric fluence matches that
of the Band function. This is a natural choice for setting the normalization,
because the photon fluence is used in Eq.\ref{eqn:nunorm}.

It is quite clear that approximation A overestimates the contribution
from high energy photons and underestimates the contribution from low energy
photons. The reason why this occurs is because the break energy of the broken
power law has been forced to match that of the Band function. 

It may be desirable (e.g. to simplify calculations) to have an alternative
broken power law approximation. Also we will use this new approximation,
\textit{B}, to illustrate the deficiencies of approximation A. A
better choice for approximating the Band function is to require that the
asymptotic behavior (for very large and very small $E_\gamma$) of the broken
power law matches the Band function, while leaving the break energy for the
broken power law a free parameter:
\begin{equation}
  \label{eqn:mybandapprox}
  \frac{dN_\gamma}{dE_\gamma}^{\mathrm{B}} = A_\gamma \left\{
  \begin{array}{lr}
    \left(\frac{E_\gamma}{100 keV}\right)^{\alpha_\gamma}  & E_\gamma < \bar{\epsilon}_\gamma \\
    \left(\frac{E_\gamma}{100 keV}\right)^{\beta_\gamma} 
\left[ \frac{(\alpha_\gamma - \beta_\gamma)}{2+\alpha_\gamma}
 \frac{E_{\mathrm{pk}}}{100 \mathrm{keV}} 
 \right]^{\alpha_\gamma-\beta_\gamma} e^{\beta_\gamma-\alpha_\gamma}
  & E_\gamma \ge \bar{\epsilon}_\gamma
  \end{array}
\right. .
\end{equation}
The value of the \textit{effective} break energy $\bar{\epsilon}_\gamma$ is
given by the energy at which the two branches of the broken power law are
equal to each other:
\begin{equation}
\left( \frac{\bar{\epsilon}_\gamma}{100 keV} \right)^{\alpha_\gamma} = 
\left( \frac{\bar{\epsilon}_\gamma}{100 keV} \right)^{\beta_\gamma} 
\left[ \frac{\alpha_\gamma - \beta_\gamma}{2+\alpha_\gamma} 
\frac{E_{\mathrm{pk}}}{100 \mathrm{keV}} \right]^{\alpha_\gamma -
  \beta_\gamma} e^{\beta_\gamma - \alpha_\gamma}
\end{equation}
Which results in an effective break energy that is independent of the spectral
indices:
\begin{equation}
\bar{\epsilon}_\gamma = \frac{{\epsilon}^{\mathrm{b}}_\gamma}{e},
\end{equation}
where $e$ is Euler's number. Figure \ref{fig:bandA} also shows approximation B
with the same parameters as before.

\section{Effect of expected number of events}
\label{sec:eventrates}

Given a neutrino spectrum $dN_\nu/dE_\nu$, the expected number of events in a
neutrino telescope is: 

\begin{equation}
N_{\mathrm{evt}} = A^{\mu} \int^{E^{\mathrm{max}}}_{E_\mu^{\mathrm{min}}} dE
P_\mu(E_\nu
;E^{\mathrm{min}}_\mu) S(E_\nu,\theta) \frac{dN_\nu}{dE_\nu},
\end{equation}
where $A^{\mu}$ is the muon effective area (1~km$^2$ for IceCube/KM3NET),
$P_\mu(E_\nu;E^{\mathrm{min}}_\mu)$ is the probability of a neutrino of energy
$E_\nu$ to produce a muon with energy equal or greater than the neutrino
telescope threshold $E^{\mathrm{min}}_\mu$ (we assume 100 GeV) and
$S(E_\nu,\theta)$ is Earth's attenuation.The probability
$P_\mu(E_\nu;E^{\mathrm{min}}_\mu)$ is given by:
\begin{equation}
P_\mu(E_\nu;E^{\mathrm{min}}_\mu) = N_A \sigma_{\mathrm{cc}}(E_\nu) <R_\mu(E_\nu;E_\mu^{
\mathrm{min}})>,
\end{equation}
where $<R_\mu(E_\nu;E_\mu^{\mathrm{min}})>$ is the average muon range given a
neutrino energy $E_\nu$ and a muon threshold $E_\mu^{\mathrm{min}}$. Earth's
attenuation factor is given by:
\begin{equation}
S(E_\nu,\theta) = e^{-z(\theta) N_A \sigma_T(E_\nu)},
\end{equation}
where $z(\theta)$ is Earth's column density as a function of angle and
$\sigma_T$  is the total $\nu$-matter crossection.

We have calculated the expected number of events for a km$^3$ detector. We
have used CTEQ5 for the neutrino-matter cross-section
\citep{2000EPJC...12..375L}. We follow the calculation by
\citep{1991PhRvD..44..3543L} of average muon range. The Earth column density
is taken from the Preliminary Earth Reference Model
\citep{1981PEPI...25..297D}.

\begin{figure}
\begin{center}
\includegraphics[width=0.45\textwidth]{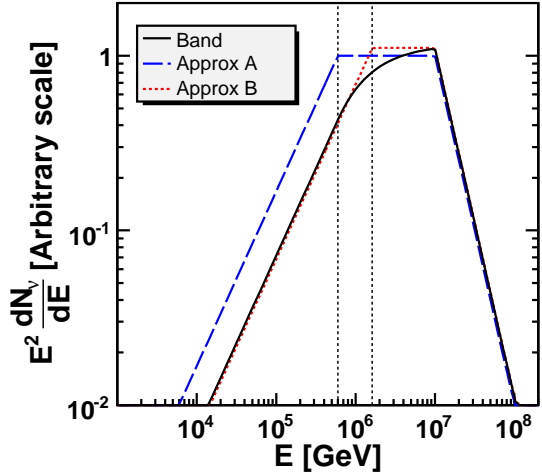}
\caption{\label{fig:NuFlux} Neutrino spectra corresponding to the Band
  function, approximation A and B with the parameters used in
  Fig.\ref{fig:bandA}. The normalization $A_\nu$ was chosen arbitrarily, but
  for approximations A and B the neutrino fluence, ignoring synchrotron
  radiation, is the same for all three spectra. The neutrino break energy is
  5.98$\times 10^{14}$~eV for the neutrino spectra derived from the Band
  funcition and for the one derived from approximation A and 1.62$\times
  10^{15}$~eV for the one derived from approximation B. The synchrotron energy
  loss has been arbitrarily set to 10$^{16}$~eV.}
\end{center}
\end{figure}

Using approximations A and B and the neutrino spectrum derived from the Band
function, we have studied a GRB with a photon break energy $\epsilon^b_\gamma
= 300$~keV, located at a redshift z=1 and with Lorentz bulk boost
$\Gamma=300$. We also set the spectral indices to $\alpha_\gamma=-2$ and
$\beta_\gamma=-1$. We have normalized the neutrino  fluence of all three
spectra to the same (arbitrary) value as described in section \S
\ref{sec:photopion}. For this GRB the effective break energy is
$\bar{\epsilon}_\gamma=110$~keV. The neutrino break energy for approximation A
is 5.98$\times 10^{14}$~eV and for approximation B it's 1.63$\times
10^{15}$~eV. For all cases we have fixed the synchrotron energy break at
10$^{16}$~eV. Figure \ref{fig:NuFlux} shows the three neutrino spectra. Again
it is clear that approximation A is inadequate because it overestimates the
contribution of low energy neutrinos.

\begin{figure}
\begin{center}
\includegraphics[width=0.45\textwidth]{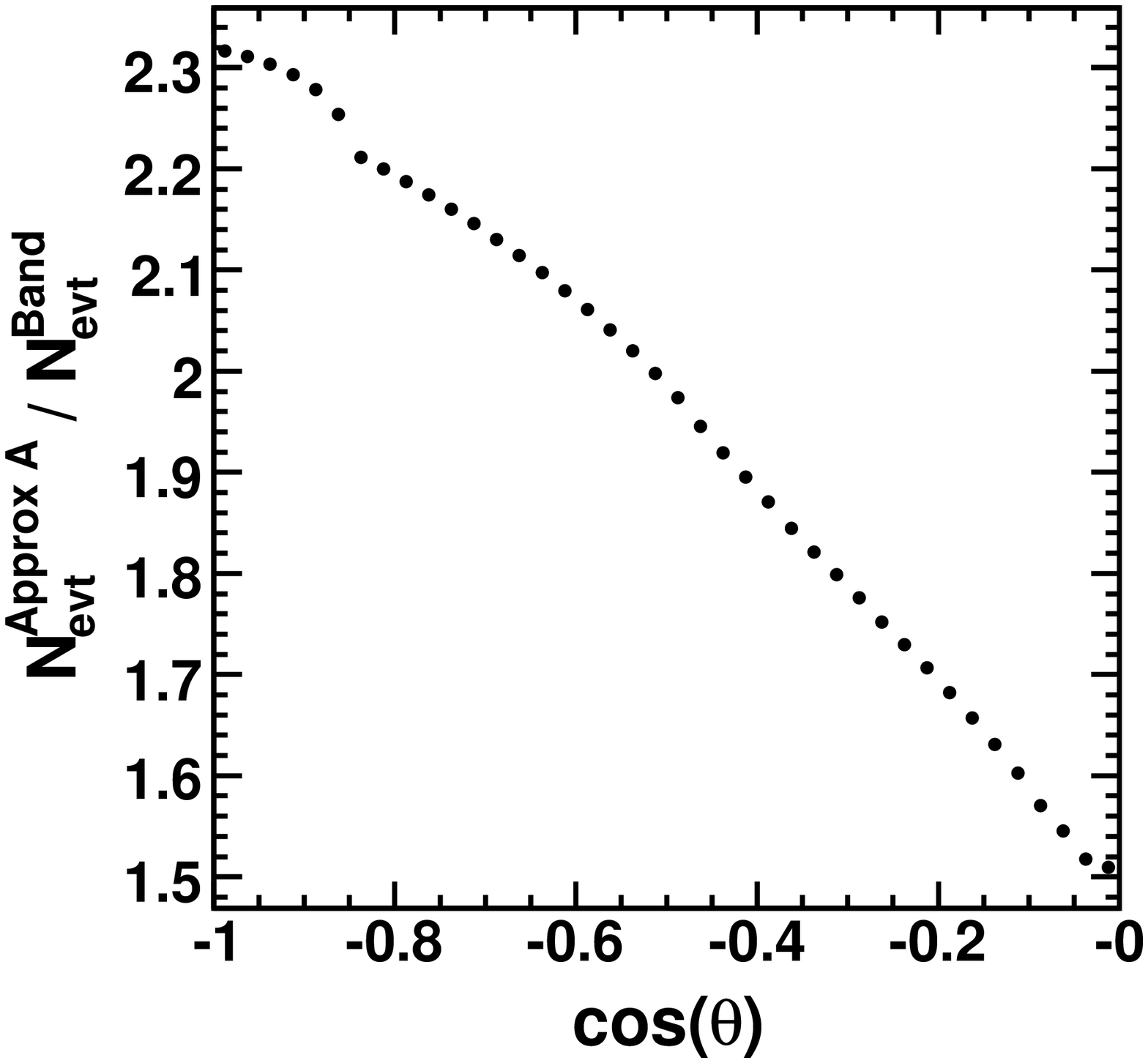}
\includegraphics[width=0.45\textwidth]{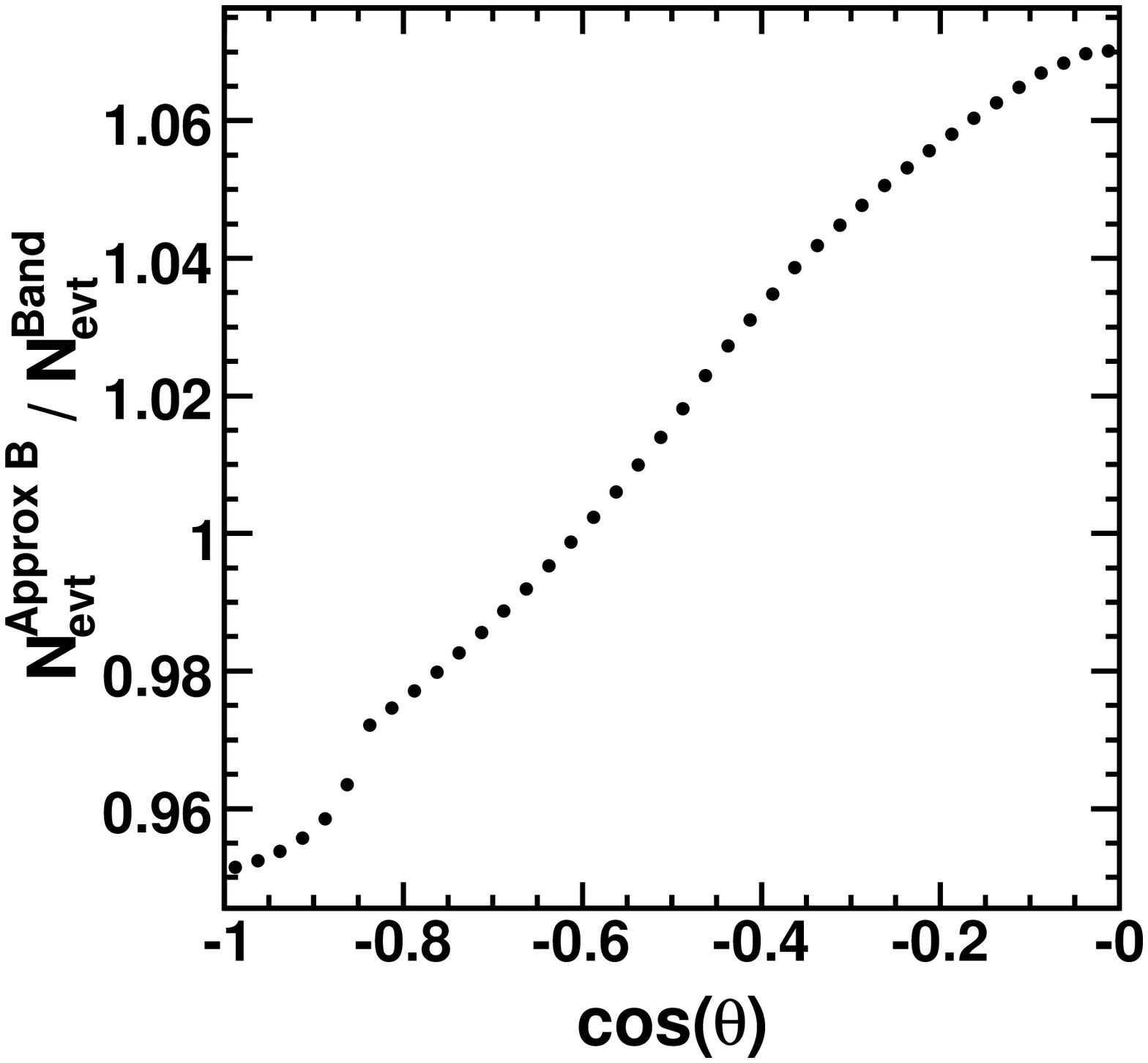}
\caption{\label{fig:ratio} Left: Ratio of expected number of events for
  approximation A to the neutrino spectrum derived from the Band function as a
  function of angle. The parameters chosen for the test GRB match those of
  Fig.\ref{fig:bandA} and \ref{fig:NuFlux}. For $\cos(\theta)=-1$ neutrinos go
  across Earth, while for $\cos(\theta)=0$ neutrinos glance Earth along the
  horizon. The structure in the plot is correlated with the density profile of
  Earth as a function of depth, e.g. Earth's core effect is seen between
  $\cos(\theta)$ -1 and -0.85. Right: Ratio of expected number of events for
  approximation B to the spectrum derived using the Band function.}
\end{center}
\end{figure}

Figure \ref{fig:ratio} the ratio approximation A to B of expected number of
events as a function $\cos(\theta)$ for this example GRB. For all GRB
locations in the sky we see that approximation A overestimates the  expected
number of events. For steeper angles the ratio is larger, because for
approximation B the characteristic neutrino energy is higher and therefore
Earth's attenuation is higher.

\section{Discussion}
\label{sec:discussion}

We have shown that the usual choice to describe the photon spectra in the
calculation of neutrino fluxes from GRBs overestimates the contribution of
high energy photons (and therefore contribution of low energy neutrinos is
overestimated). This results in a higher exepected event rate by a factor of
$\approx$~2 for all models that make this assumption. The actual value of the
overestimation of the expected number of events depends on the matter column
depth that neutrinos must cross through Earth in the direction of the
GRB. Also we have shown that the typical neutrino energy is
$\approx$~10$^{15}$~eV. The characteristic energy of neutrinos is a factor of
$e$ larger than the values obtained by \citet{2004APh....20..429G} and a
factor of 10 than the average value used by \citet{1997PhRvL..78.2292W}. For
back of the envelope calculations we provide a new approximation to the Band
function that is adequate for the computation of neutrino spectra.

\citet{2005PhRvL..95..181101K} have discussed the effects of muon and pion
energy losses leading to neutrino flavor flux ratios (at Earth) different than
1:1:1. Previous AMANDA searches for neutrinos
\citep{2007ApJ...664..397A,2007ApJ...InPress}, based on Waxman-Bahcall-like
assumptions have argued that this effect is not important experimentally
because the typical energy of the neutrinos detected is close to the neutrino
break energy. Because, as we have shown, the \textit{effective} neutrino break
energy is close to the typical syncrhotron energy break, the effects described
by \citet{2005PhRvL..95..181101K} must indeed be relevant.

\acknowledgments
I.T and M.V.D. were supported in part by NSF ANT-0554699.


\begin{thebibliography}{}
 
\bibitem[Achterberg et al(2007a)]{2007ApJ...664..397A}
Achterberg A. et al. 
\newblock{ { \em\apj} 664:397, 2007}

\bibitem[Achterberg et al(2007b)]{2007ApJ...InPress}
Achterberg A. et al. 
\newblock{ { \em \apj} In press, 2007}

\bibitem[Achterberg et al(2007c)]{2007..IceCube..ICRC}
Achterberg A. et al. 
\newblock {Proc. of 30$^{th}$ Int. Cosmic Ray Conf. 2007. arXiv:0711.0353}

\bibitem[Band et al(1993)]{1993ApJ...413..281B}
Band~D. et~al.
\newblock{ {\em \apj}, 413:281, 1993.}

\bibitem[Dziewonski \& Anderson(1981)]{1981PEPI...25..297D}
Dziewonski~A.~M. and Anderson~D.~L.
\newblock{ {\em Phys. Earth Plan. Int.} 25:297, 1981}

\bibitem[Guetta et al(2004)]{2004APh....20..429G}
Guetta~D. et~al.
\newblock{ {\em Astroparticle Physics}, 20:429, 2004.}

\bibitem[Halzen \& Hooper(1999)]{1999ApJ...527L..93H}
Halzen~F. and Hooper~D.~W.
\newblock{ {\em \apjl}, 527:L93, 1999.}

\bibitem[Kashti \& Waxman(2005)]{2005PhRvL..95..181101K}
Kashti~T. and Waxman~E.
\newblock{ {\em \prl}, 95:181101, 2005.}

\bibitem[Kappes et al(2007)]{2007..KM3NET..Faro}
Kappes~A. et~al. 
\newblock{ {Proc 6$^{th}$ Int Workshop on New Worlds in
  Astropart. Phys. 2007. arXiv:0711.0563 }}

\bibitem[Lai et al(2000)]{2000EPJC...12..375L}
Lai~H.~L. et al
\newblock{ {\em European Physical Journal C}, 12:375, 2000.}

\bibitem[Lipari \& Stanev(1991)]{1991PhRvD..44..3543L}
Lipari~P. and Stanev~T.
\newblock{ {\em \prd} 44:3543, 1991}

\bibitem[Meszaros(2006)]{2006RPPh...69.2259M}
Meszaros~P.
\newblock{ {\em Reports of Progress in Physics},69:2259, 2006 }

\bibitem[Murase \& Nagataki(2006)]{2006PhRvD..73f3002M}
Murase~K. and Nagataki~S.
\newblock {\em \prd}, 73:063002, 2006.

\bibitem[Vietri(1995)]{1995ApJ...453..883V}
Vietri~M.
\newblock {\em \apj}, 453:883, 1995.

\bibitem[Waxman(1995)]{1995PhRvL..75..386W}
Waxman~E.
\newblock {\em \prl}, 75:386, 1995.

\bibitem[Waxman \& Bahcall(1997)]{1997PhRvL..78.2292W}
Waxman~E. and Bahcall~J.
\newblock {\em \prl}, 78:2292, 1997.

\end{thebibliography}
\end{document}